\documentstyle[aps,prl,multicol,epsf]{revtex}

\def\Tr{\mbox{Tr}}

\newcommand{\bleq}{\ifpreprintsty
                   \else
                   \end{multicols}\vspace*{-3.5ex}{\tiny
                   \noindent\begin{tabular}[t]{c|}
                   \parbox{0.493\hsize}{~} \\ \hline \end{tabular}}
                   \fi}
\newcommand{\eleq}{\ifpreprintsty
                   \else
                   {\tiny\hspace*{\fill}\begin{tabular}[t]{|c}\hline
                    \parbox{0.49\hsize}{~} \\
                    \end{tabular}}\vspace*{-2.5ex}\begin{multicols}{2}
                    \fi}
\newcommand{\bcols}{\ifpreprintsty\else\begin{multicols}{2}\fi}
\newcommand{\ecols}{\ifpreprintsty\else\end{multicols}\fi}

\begin{document}
\bibliographystyle{prsty}
\title{Keldysh and Doi--Peliti Techniques for out--of--Equilibrium
Systems.}  
\draft
\author{ Alex Kamenev}
\address{ Department of Physics, Technion, Haifa 32000, Israel.
  \\
  {}~{\rm (\today)}~
  \medskip \\
  \parbox{14cm} 
    {\rm Lecture notes presented at Windsor NATO school on 
 "Field Theory of Strongly Correlated Fermions and
   Bosons in Low-Dimensional Disordered Systems" (August 2001). 
The purpose of these lectures is to give a brief modern introduction to  
Keldysh non--equilibrium field theory and its classical analog --  
Doi--Peliti technique. The special emphasis is put on stressing the analogy between  the two approaches. 
    \smallskip\\ }\bigskip \\ }
\maketitle

\section{Quantum Keldysh formalism}

1. {\em A bit of history.}
The name of the technique takes its origin  in  1964 paper 
of L.~V.~Keldysh \cite{Keldysh65}. Among the earlier closely related 
approaches, one should mention J.~Schwinger \cite{Schwinger61} and 
R.~P.~Feynman and F.~L.~Vernon \cite{Feynman63}.  Classical 
counterparts of the Keldysh technique are extremely useful and interesting 
on their own. I shall mention  Wyld diagrammatic technique
\cite{Wyld}, which is the basis of the modern theory of turbulence; 
Martin--Siggia--Rose--DeDominics method \cite{MSR} for 
stochastic systems; and  Doi--Peliti technique 
\cite{Doi,Peliti} for reaction--diffusion systems and cellular automata. 
Despite the unique power 
and wide popularity of the technique its pedagogical presentations are 
rare. One may find some  in  books 
\cite{Landau,Mahan} or in a review \cite{Rammer}. 
This lecture is not an alternative for the existing literature. 
Using the simplest possible example, I try to explain 
the general structure of the theory, its relation to other 
approaches, its potential for novel applications and open problems. 
Technicalities of the numerous practical applications may be found 
in the literature and  left outside the scope of the lecture.  
I tried hard, however, to mention possible traps and silent points 
which are rarely exposed explicitly.

\vskip 1cm

2. {\em Motivation}. The Keldysh formulation of the many--body theory is 
useful in the following cases:
\begin{itemize}
\item For treatment of  systems not in the thermal equilibrium 
\cite{Keldysh65}. 
\item For calculation of the full counting statistics of a
quantum mechanical observable (as opposed to an average value or 
correlators) \cite{Levitov,Nazarov}.
\item As an alternative to the replica and supersymmetry  methods in the 
theory of disordered and glassy systems 
\cite{Sompolinsky,Feigelman,Kamenev99,Shamon99,Altland00}.
\item For equilibrium problems where Matsubara analytical continuation may 
prove to be sophisticated \cite{Rammer}.

\end{itemize}

\vskip 1cm

\begin{figure}
\vglue 0cm
\hspace{0.01\hsize}
\epsfxsize=1.0\hsize
\epsffile{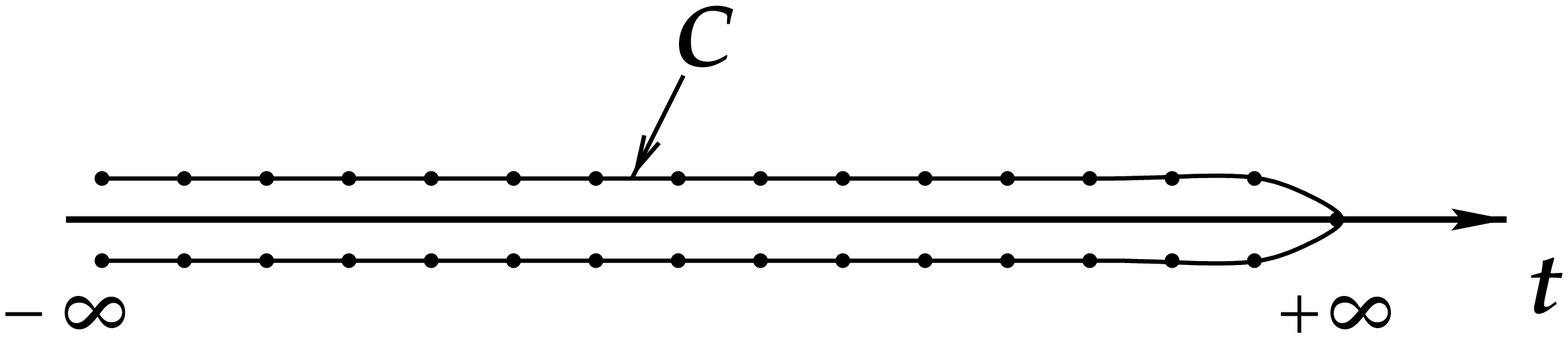}
\refstepcounter{figure} \label{contour}
{\small FIG.\ \ref{contour}
The closed Keldysh time contour ${\cal C}$. 
Dots on the forward  and the backward branches of the contour 
denote the discretized time points.
  \par}
\end{figure}

3. {\em Closed time contour}.
The standard construction of the zero temperature (or equilibrium) 
many--body theory (see e.g. \cite{AGD,Mahan}) involves the procedure 
of adiabatic switching  interactions on at distant past and then switching
it  off at distant future. The crucial assumption here is that starting 
from the ground (or equilibrium) state of the system at $t=-\infty$ one 
arrives to the same state at $t=+\infty$ (acquiring some 
phase factor along the way). This is clearly not the case out of the 
equilibrium. Starting from some arbitrary distribution and then switching 
interactions on and off, one is going to find the system in some unpredictable 
state. The latter depends, in general, on the peculiarities of the switching 
procedure. 
The absence of knowledge about the finite state spoils completely  the entire
construction \cite{AGD,Mahan}.  One would like thus to build a theory which 
avoids reference to the state at $t=+\infty$. We still need to know a finite 
state (since we calculate traces). 
The Schwinger suggestion is to make the finite state 
to be exactly the initial one.  
The central idea  is to let a quantum system to evolve 
first in the forward direction in time and then to "rewind" its 
evolution back, playing the  movie in the backward direction.
One  ends up thus with the need to construct a theory with the time evolution
along the two--branch contour depicted on Fig.~\ref{contour}.
Then, no matter what was the state at $t=+\infty$, after the backward evolution 
it will return back to the known initial state. In this construction there is no 
switching of interactions in the future. Both switchings on and off take place 
in the past: on -- on the forward branch and off -- on the backward.
How to construct such a theory and how to use it -- is the subject of  this chapter.

\vskip 1cm

4. {\em Field theory}. To be specific, let me consider the simplest 
possible many--body system. It consists of bosons living in a single  
quantum state with the energy $\omega_0$: 
\begin{equation}
H = \omega_0 a^{\dagger} a\, ;
                                         \label{Ham}
\end{equation}
here $a^{\dagger}$ and $a$ are bosonic creation and annihilation 
operators. Let us consider the "partition function" defined as 
\begin{equation} 
Z=\Tr\{\rho_0  U_{\cal C}\} / \Tr\{\rho_0 \} \, , 
                                                              \label{e1}
\end{equation}
where $U_{\cal C}$ is the evolution operator along the closed contour
${\cal C}$. If we assume (as we shall do for a while) 
that all the external fields are exactly the same on the forward and 
backward branches of the contour, then $U_{\cal C}=1$ and therefore 
$Z=1$. In Eq.~(\ref{e1})
$\rho_0=\rho(H)$ is some 
\footnote{To accommodate the equilibrium initial density matrix, 
$\rho_0 = \exp\{-\beta (H-\mu N)\}$, it is sometimes
suggested to add an imaginary (vertical) part to the Keldysh contour 
at $t=-\infty$. To my opinion this procedure only obscures the structure 
of the theory. Indeed, the technique is not restricted in any way to the 
initial distribution  being  the equilibrium one. 
Therefore it is much preferred to keep the structure general and not commit 
oneself with unnecessary assumptions.}
density matrix operator defined at $t=-\infty$. 
In our example $\Tr\{\rho_0 \}=[1-\rho(\omega_0)]^{-1}$. 
An important point is that in general 
$\Tr\{\rho_0 \}$ is an  interaction-- and disorder--{\em independent} 
constant. Indeed both interactions and disorder are supposed to be switched on 
(and off) on the forward (backward) parts of the contour sometime after (before)
$t=-\infty$. I shall therefore frequently omit this constant or refer to it 
as being unity -- it never causes a confusion. 

The next step is to divide the ${\cal C}$ contour into $2N+1$ time
steps of the length $\delta_t$, 
such that $t_1=t_{2N+1}=-\infty$ and $t_{N+1}=+\infty$ as shown on
Fig.~\ref{contour}.  Following the
standard route \cite{Negele}, we obtain the coherent state functional
integral, by introducing a resolution of unity at each time
step. Taking the $N\to \infty$ limit we obtain for
the partition function  
\begin{equation} 
Z= {1\over \Tr\{\rho_0\} }
\int\! {\cal D}\bar\phi \phi\,  \exp\{iS[\bar\phi(t),\phi(t)] \}\, ,
                                                              \label{e2}
\end{equation}
where the action is given by 
\begin{equation} 
S[\bar\phi(t),\phi(t)] = \int\limits_{\cal C}\! dt\, 
\bar\phi(t) {\cal D}^{-1}\phi(t)\, ,
                                                              \label{e3}
\end{equation}
and  ${\cal D}^{-1}=[i\partial_t - \omega_0]$. It is important to 
remember that this continuous notation is only a short way to represent 
the $2N \times 2N$ matrix ${\cal D}^{-1}_{ij}$
\begin{equation}
  \label{Dmatrix}
  -i{\cal D}^{-1}_{ij} =\left[\matrix{
1 & 0 &  & & & &&(-1+\delta_t i\omega_0)\rho(\omega_0)  \cr
-1+\delta_t i\omega_0 & 1 & 0 &  & & &&  \cr
 &-1+\delta_t i\omega_0 & 1 & 0 &  & & & &&\cr
 &  & -1+\delta_t i\omega_0 & 1 & 0 &  & & && \cr
& & \hspace{.4cm} \ddots& \hspace{.4cm}\ddots &\hspace{.4cm}\ddots && &  \cr
&&&&&&&\cr
&&&&&&&\cr
&&&&&&&\cr
& &&& & -1 - \delta_t i\omega_0& 1& 0\cr 
& &&& & & -1 - \delta_t i \omega_0 & 1
}\right]\, ,
\end{equation}
The change of sign $+\delta_t \to -\delta_t$ takes place in the center 
of the matrix at the point where the time axis changes its direction.
It is straightforward to evaluate the determinant of such  matrix 
\begin{equation}
\mbox{det} \big[ -i{\cal D}^{-1} \big] =
1+(-1)^{2N+1}\rho(\omega_0)(-1+\delta_t i\omega_0)^N  (-1-\delta_t i\omega_0)^N  
\to 1- \rho(\omega_0) \, ; 
                                                \label{determinant}
\end{equation}
Recalling that $\rho(\omega_0) <1$, one may convince himself that 
all eigenvalues of the matrix $-i{\cal D}^{-1}_{ij}$ have positive real part. 
This fact secures 
convergence of the functional integral. Performing the integration, one 
not--surprisingly obtains for the partition function 
\begin{equation}
Z= \frac{ \mbox{det}^{-1}\big[ -i{\cal D}^{-1} \big]}{\Tr\{\rho_0 \} } = 1\, .
                                                \label{unity}
\end{equation}
Seemingly absent in the continious notations -- the 
upper right element of the matrix is extremely important to obtain the correct 
normalization. It is not less important to find the proper correlation functions.

We divide next the bosonic field $\phi(t)$ onto the two components 
$\phi_f(t)$ and $\phi_b(t)$ which reside on the forward and the 
backward parts of the time contour correspondingly. 
The action may be now rewritten as $S[\phi_f] -S[\phi_b]$, with the time 
integration running along the real axis  
(the minus sign comes from reversing direction of the time integration on 
the backward part of the contour). 

\vskip 1cm

5. {\em Correlation functions.}
Now comes the most unpleasant part: to invert the matrix and find the 
correlation (or Green) functions. We may save ourselves a lot of efforts by 
recalling that the functional integral always produces the time--ordered 
Green functions \cite{Negele} and that the time arguments on the backward branch are 
always {\em after} those on the forward. This way one finds
\footnote{The more familiar definition of the Green functions is 
${\cal D}=-i\langle \phi\bar\phi\rangle/Z$, since in our case $Z=1$ --
we are doing OK.}  
\begin{eqnarray}
-i\langle \phi_f(t)\bar\phi_f(t')\rangle &=& {\cal D}^T(t,t')=
\theta(t-t'){\cal D}^>(t,t') + \theta(t'-t){\cal D}^<(t,t') \, ; 
\nonumber \\
-i\langle \phi_f(t)\bar\phi_b(t')\rangle &=& 
{\cal D}^<(t,t') = \left(-ine^{-i\omega_0(t-t')}\right) \, ;
\nonumber \\
-i\langle \phi_b(t)\bar\phi_f(t')\rangle &=& 
{\cal D}^>(t,t') = \left(-i(n+1)e^{-i\omega_0(t-t')} \right)\, ;
\nonumber \\
-i\langle \phi_b(t)\bar\phi_f(t')\rangle &=& 
{\cal D}^{\tilde T}(t,t') = 
\theta(t'-t){\cal D}^>(t,t') + \theta(t-t'){\cal D}^<(t,t') \, , 
                                         \label{corr}
\end{eqnarray}
where the angular brackets are understood as the functional 
integration with the action given above. The structure is general,
but ${\cal D}^{>} = -i\langle |a(t) a^\dagger(t')|\rangle$ and 
${\cal D}^{<}= -i\langle |a^\dagger(t') a(t)|\rangle$ are given for 
orientation in brackets for our simple example 
($n=\rho(\omega_0)/(1-\rho(\omega_0))$ is the bosonic 
occupation number). 

It may be surprising 
to see the non--zero off--diagonal correlators 
$\langle \phi_f\bar\phi_b\rangle$ and 
$\langle \phi_b\bar\phi_f\rangle$, while the  action seems
to be perfectly diagonal in the $f-b$ space. The trick is exactly in 
the presence of the upper--right element of the discretized matrix, hidden 
in the continuous representation. Notice, that the presence 
of the off--diagonal elements does not contradict to the continuous notations. 
Indeed, 
$[i\partial_t - \omega_0]{\cal D}^{>,<} = 0$, while 
$[i\partial_t - \omega_0]{\cal D}^{T,\tilde T} = \pm \delta(t-t')$. 
Therefore in the obvious $2\times 2$ matrix notations 
${\cal D}^{-1}{\cal D} = 1$,  as it should be.
The point is that the  $[i\partial_t - \omega_0]^{-1}$ 
operator is not uniquely defined. The boundary conditions must be 
specified and the upper right element does exactly this.

Obviously not all four Green functions
defined above are independent. Indeed, an inspection shows that 
\begin{equation}
{\cal D}^T + {\cal D}^{\tilde T} = {\cal D}^> + {\cal D}^< \,. 
                                                \label{relation}
\end{equation}
One would therefore like to perform a linear transformation of 
the fields to benefit  explicitly from this relation. This is achieved 
by the  Keldysh rotation.

\vskip 1cm

5. {\em Keldysh rotation.}
Let us define the new fields as 
\begin{equation}
\phi_{cl}(t) = {1\over 2}(\phi_f(t) + \phi_b(t)) \,; \,\,\,\,\,\,\,\,\,
\phi_{q}(t) =  {1\over 2}(\phi_f(t) - \phi_b(t))\, 
                                                \label{rotation}
\end{equation}
with the analogous transformation for the conjugated fields. 
The integration measure changes by unessential multiplicative constant.
The subscripts 
{\em cl} and {\em q} stand for {\em classical} and {\em quantum} components 
of the fields correspondingly. The rational for such notations will become 
clear shortly. First, a little algebra shows that
\footnote{For fermions it is more convenient to transform 
$\psi$ and $\bar\psi$ in the different manner (it is allowed since 
$\psi$ and $\bar\psi$ are not actually complex conjugated, but rather 
notations). As a result, the standard matrix form of the 
fermionic propagator is 
$$
\left(\begin{array}{cc}
{\cal G}^R(t,t') & {\cal G}^{K}(t,t') \\
0 &   {\cal G}^{A}(t,t') 
\end{array}\right)\, .
$$
In this notation fermion--boson interactions take an especially simple 
form $S_{int} \sim \phi_\alpha \bar\psi \gamma^{\alpha} \psi$,
with $\gamma^{cl} = 1$ and $\gamma^{q}=\sigma_1$ 
\cite{Kamenev99}. Since fermions are never a classical field, this form 
of the fermionic propagator is not related to our subsequent discussion.}
\begin{equation}
-i\langle \phi_\alpha(t)\bar\phi_\beta(t')\rangle=
{1\over 2} \left(\begin{array}{cc}
{\cal D}^K(t,t') & {\cal D}^{R}(t,t') \\
{\cal D}^{A}(t,t') & 0
\end{array}\right)\, ,
                                                \label{Green}
\end{equation}
where $\alpha, \beta = cl,q$. Here indexes $R,A,K$ stay for 
{\em retarded, advanced} and {\em Keldysh} components of the Green 
function. These three are the main objects of the Keldysh technique. 
They are defined as
\begin{eqnarray}
{\cal D}^{R} &=& {\cal D}^{T} - {\cal D}^{<}= 
\theta(t-t')({\cal D}^{>} - {\cal D}^{<})\, ;   \nonumber \\
{\cal D}^{A} &=& {\cal D}^{T} - {\cal D}^{>}= 
\theta(t'-t)({\cal D}^{<} - {\cal D}^{>})\, ;   \nonumber \\
{\cal D}^{K} &=& {\cal D}^{>} + {\cal D}^{<}\, .
                                         \label{Greens}
\end{eqnarray}
Notice that the retarded (advanced) component is lower (upper) triangular 
matrix in the time space. Moreover, our regularization of the functional 
integral guarantees that these functions never appear at exactly coinciding
arguments, removing thus uncertainty about the value 
$\theta(t-t) = \theta(0)$.
Fields $\bar\phi$  always appear at one time step $\delta_t$
{\em after} $\phi$ fields on the Keldysh contour. As a result, 
$\bar\phi_f(t)$ is after $\phi_f(t)$ and $\bar\phi_b(t)$ is before $\phi_b(t)$. 
Employing Eqs.~(\ref{corr}) and (\ref{Greens}), 
one may see   that with this convention 
${\cal D}^R(t, t) \equiv {\cal D}^R(t, t+\delta_t) =0$ and 
${\cal D}^A(t,t) \equiv {\cal D}^A(t+\delta_t,t) =0$. 
In the continuous notations it is equivalent to the choice of the 
regularization, where $\theta(0) = 0$.
As a result,  for any number 
\footnote{This is true for a single function as well. For example, if 
${\cal D}^{R}=(Dq^2-i\omega)^{-1}$, one may worry that 
$\int d\omega (Dq^2-i\omega)^{-1}$ is divergent. This is not the case,
however, since this integral is simply ${\cal D}^{R}(t,t)=0$.} 
of retarded or advanced functions one has 
\begin{eqnarray}
\Tr\{ {\cal D}^{R} {\cal D}^{R} \ldots {\cal D}^{R} \} &=& 0 \, ;   
\nonumber \\
\Tr\{ {\cal D}^{A} {\cal D}^{A} \ldots {\cal D}^{A} \} &=& 0 \, ,
                                         \label{traces}
\end{eqnarray}
where the traces are understood as time integrations. If a problem 
is translationally invariant in time -- the same statement has the simple 
reinterpretation
\begin{equation}
\int\limits_{-\infty}^{\infty}\! d\omega\,  
{\cal D}^{R}(\omega) {\cal D}^{R}(\omega+\Omega_1) \ldots 
{\cal D}^{R}(\omega+\Omega_n) =0\, ,
                                             \label{traces1}
\end{equation}
which is an obvious consequence of retarded Green function being an 
analytic function in the entire upper half plane of complex energy. 
From Eq.~(\ref{Green}) it is obvious that 
\begin{equation}
{\cal D}^{A} = \left[ {\cal D}^R \right]^{\dagger}\, ,
                                              \label{conj}
\end{equation}
where Hermit conjugation includes transposition of the time arguments. 
 
In our simple example 
${\cal D}^{>} = -i(n+1)e^{-i\omega_0(t-t')}$ and 
${\cal D}^{<} = -ine^{-i\omega_0(t-t')}$, where 
$n=n(\omega_0)= \rho(\omega_0)/(1-\rho(\omega_0))$ is the bosonic 
occupation number (since the system is non--interacting the initial 
distribution function does not evolve).  As a result 
\begin{eqnarray}
{\cal D}^{R}(t,t') &=& 
-i\theta(t-t') e^{-i\omega_0(t-t')} \, ;   \nonumber \\
{\cal D}^{A}(t,t') &=& 
i\theta(t'-t)e^{-i\omega_0(t-t')}\, ;   \nonumber \\
{\cal D}^{K}(t,t') &=& -i (2n(\omega_0)+1)e^{-i\omega_0(t-t')} \, .
                                         \label{Greens1}
\end{eqnarray}
Notice that the retarded and advanced components are independent on the 
distribution function, whereas the Keldysh component does depend on it.
This statement is much more general than our toy example. Going to the 
frequency representation, one finds that 
${\cal D}^{K}(\omega) = -2\pi i (2n(\omega_0)+1)\delta(\omega-\omega_0) = 
-2\pi i(2n(\omega)+1)\delta(\omega-\omega_0)$. 
In the particular case of the thermal  equilibrium, one has  
\begin{equation}
{\cal D}^{K}(\omega) =\mbox{coth}{\beta\omega \over 2}
\left( {\cal D}^{R}(\omega) - {\cal D}^{A}(\omega) \right)\, .
                                         \label{fdt}
\end{equation}
The last equation constitutes the statement
\footnote{The analogous statement for  fermions reads as 
$$
{\cal G}^{K}(\epsilon) =\mbox{tanh}{\beta\epsilon \over 2}
\left( {\cal G}^{R}(\epsilon) - {\cal G}^{A}(\epsilon) \right)\, .
$$}
of the 
{\em fluctuation--dissipation theorem} (FDT). 
It implies (as we shall see below)  the rigid relation
between response and correlation functions at equilibrium.
FDT is always satisfied in 
the thermal  equilibrium even in the presence of the interactions. 
We shall see later why the interactions  do not change this 
relation. In  general case it is convenient to define a function 
(matrix in the time space) $F$ via
\begin{equation}
{\cal D}^{K} =
 {\cal D}^{R} F  - F {\cal D}^{A}  \, 
                                         \label{gfdt}
\end{equation} 
and refer to $F$ (or better its Wigner transform) 
as the distribution function. At equilibrium $F(t-t') \to 
F(\omega) =\mbox{coth}{\beta\omega \over 2} = 2n(\omega) +1$,
where $n(\omega)$ is the Planck distribution.

\vskip 1cm

6. {\em Keldysh action and causality.}
Once we have understood the structure of the  propagator in the 
$cl-q$ space, we may go back to the action. Inverting the 
matrix Eq.~(\ref{Green}),  one finds 
\footnote{Notice that the matrix in the exponent of the functional integral, 
$i{\cal D}^{-1}$,  is {\em anti}--Hermitian ! The 
convergence factors are hidden in the regularization.}
\begin{equation} 
S[\phi_{cl},\phi_q] = 2  \int\!\!\int\limits_{-\infty}^{\infty}\!\! dt dt'\, 
(\bar\phi_{cl}(t),\bar\phi_q(t)) 
\left(\begin{array}{cc}
0   & [{\cal D}^{A}]^{-1}  \\
  \left[{\cal D}^{R}\right]^{-1}  & [{\cal D}^{-1}]^K 
\end{array}\right)_{(t,t')}
\left(\begin{array}{c}
\phi_{cl}(t') \\ \phi_q(t')
\end{array}\right) \, ,
                                                              \label{action}
\end{equation}
where $[{\cal D}^{-1}]^K = [{\cal D}^{R}]^{-1} F - F [{\cal D}^{A}]^{-1}  \neq
[{\cal D}^{K}]^{-1}$. Of course, in our simple example 
$[{\cal D}^{R,A}]^{-1} = \delta(t-t')[i\partial_{t'} -\omega_0 \pm i0]$ and 
$[{\cal D}^{-1}]^K =0$ and therefore the correlator matrix  is basically 
$\delta(t-t')[i\partial_{t'} -\omega_0]\sigma_1$, as it should be. 
The point is that the structure 
of the Gaussian action given by Eq.~(\ref{action}) is absolutely generic and correctly 
encodes regularization of the functional integral. 
Since the Keldysh component carries information about the density matrix, 
there is no need any more to remember about the discreet representation. 
As we shall see below this structure
remains intact under any perturbative renormalization.  For the lack of a better 
terminology we shall refer to the robustness of this structure as the {\em causality} 
principle \cite{Altland00} (since it is the reason for the response functions being 
casual, see below). Let us devote some time to  this  structure.

Consider first the $cl-cl$ element, which is zero. 
This zero may be traced back to the relation~(\ref{relation}). 
It has, however, the  much simpler interpretation. It says 
that if one takes a purely classical field configuration, $\phi_q=0$, the action 
is zero.
But this is obvious, since in this case $\phi_f=\phi_b$ and the action on the 
forward contour cancels exactly that on the backward. 
We therefore arrive to the  simple but extremely important statement: 
\begin{equation} 
S[\phi_{cl},\phi_q=0] = 0\, .
                                                              \label{causality}
\end{equation}
From this perspective the 
relation~(\ref{relation}) is simply one of manifestations of the causality. Obviously 
Eq.~(\ref{causality}) is not restricted to the Gaussian action. Consider e.g. 
the simplest interaction action 
$S_{int} = {\lambda\over 2}\int_{\cal C}\!\! dt\, (\bar\phi \phi)^2 = 
{\lambda\over 2} \int\! dt\,  [(\bar\phi_f \phi_f)^2 - (\bar\phi_b \phi_b)^2]$. 
Notice that this is the  continuous 
notation; in discreet version $\bar\phi$ fields on the Keldysh 
contour ${\cal C}$ are taken 
at one step $\delta_t$ {\em later}
time with respect to $\phi$ fields.
Transforming to the classical--quantum 
components, one finds
\begin{equation} 
S_{int}[\phi_{cl},\phi_q] = 4 \lambda\, \Re\!\!  
\int\limits_{-\infty}^{\infty}\!\! dt\,\,  
\bar\phi_q \bar\phi_{cl}( \phi_{cl}^2  + \phi_q^2 ) \, ,
                                                              \label{int}
\end{equation} 
which obviously satisfy Eq.~(\ref{causality}).

\begin{figure}
\vglue 0cm
\hspace{0.01\hsize}
\epsfxsize=0.7\hsize
\epsffile{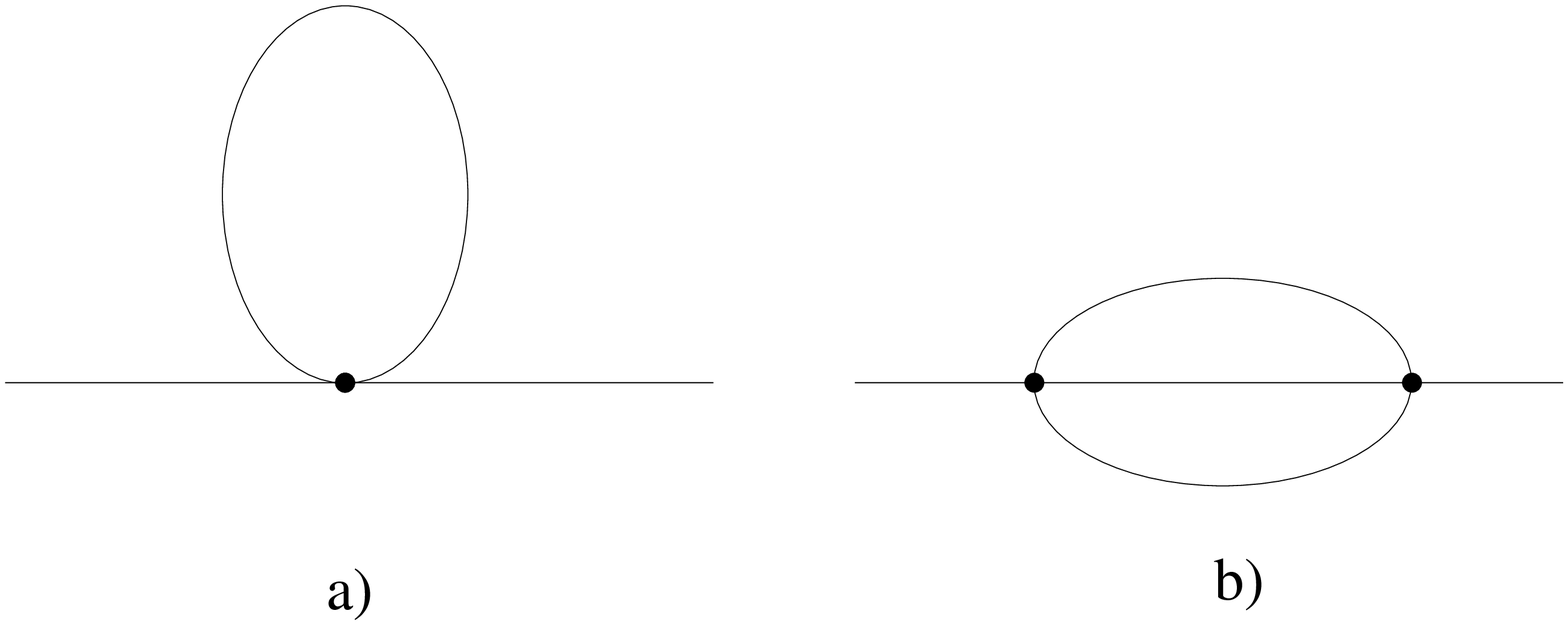}
\refstepcounter{figure} \label{phi4}
{\small FIG.\ \ref{phi4}
First and second order diagrams for the $(\bar\phi\phi)^2$ theory. 
  \par}
\end{figure}

The fact that $cl-q$ and $q-cl$ components in Eq.~(\ref{action}) 
are  advanced and retarded (means 
triangular in the time representation) is of the fundamental 
importance as well. Due to this fact Eq.~(\ref{causality}) is intact  in 
any order of the perturbation theory. Consider, for example, the first order correction 
in $\lambda$ to the $cl-cl$ element of the Gaussian action (which is zero). 
The corresponding diagram is depicted on Fig.~\ref{phi4}a. 
From Eq.~(\ref{int}) one reads out
\begin{equation} 
\delta S_{cl-cl} = 4\lambda\, \Re\!\! \int\limits_{-\infty}^{\infty}\! dt\, 
\bar\phi_{cl}(t)  {\cal D}^{R}(t,t) \phi_{cl}(t)\, . 
                                                              \label{diag1}
\end{equation} 
As was explained above, ${\cal D}^{R}(t,t) = 0$, and the  
$cl-cl$ element obviously remains zero. 
Employing Eqs.~(\ref{traces}), (\ref{traces1}) one may show that all higher  
loop corrections do not change this fact. One may say that 
Eqs.~(\ref{traces}), (\ref{traces1}) are part of  the causality principle.
To complete the prove that causality is preserved in perturbation theory, one 
has to show that $cl-q$ and $q-cl$ components remain purely advanced and retarded
correspondingly. A reader is advised to check that this is indeed the case in the 
second order, Fig.~\ref{phi4}b.

\vskip 1cm

7. {\em Saddle point equations.}
Before doing the perturbation theory we have to discuss saddle points of the action.
Eq.~(\ref{causality}) states that there are no terms in the action which 
have zero power of both $\bar\phi_q$ and $\phi_q$. The same is obviously 
true regarding 
$\delta S/ \delta \bar\phi_{cl}$ and therefore the saddle point equation
\begin{equation}
{\delta S\over \delta \bar\phi_{cl} } = 0\, 
                                             \label{sp1}
\end{equation}
may be always solved by 
\begin{equation}
\Phi_q = 0 \, .
                                             \label{cl1}
\end{equation}
The saddle point solutions are denoted by capital $\Phi$. 
One may  check that this is
indeed the case for the action  given by 
Eqs.~(\ref{action}), (\ref{int}). Under the condition Eq.~(\ref{cl1}) the 
second saddle point equation takes the form
\begin{equation}
{1\over 2}\, {\delta S\over \delta \bar\phi_{q} } = 
{\cal O}^R[\Phi_{cl}] \Phi_{cl} =0
\, ,
                                             \label{sp2}
\end{equation}
where ${\cal O}^R[\Phi_{cl}] $ is  retarded operator describing the classical 
dynamics of the field. In our example
\begin{equation}
{\cal O}^R[\Phi_{cl}]  = i\partial_t -\omega_0 +\lambda |\Phi_{cl}|^2\, 
                                             \label{GP}
\end{equation}
and the saddle point equation~(\ref{sp2}) is the time--dependent "Gross--Pitaevskii" 
equation. We have arrived to the conclusion that among possible saddle point 
solutions of the Keldysh action, there are always some  with zero quantum 
component and the classical component obeying the classical equation of motion.
We shall call such saddle points {\em classical} or {\em casual} (since 
perturbation theory on top of a classical solution preserves the 
causality structure). 
Notice, that the classical equation of motion is  result of   variation
over the {\em quantum}  component (in the limit where the later is zero).

In view of Eqs.~(\ref{causality}) and (\ref{cl1}), the action 
on any classical solution is zero
\footnote{This statement is a close relative of the Parisi--Efetov--Wegner theorem in 
supersymmetry or the fact that a replica symmetric saddle point results in zero action
in the replica limit.}. Recall now that we do the theory for $Z=1$ -- this is 
the exact relation. To reproduce it in the saddle point approximation one has to show 
that the fluctuation determinant on top of the classical saddle point is unity.
It may be a formidable task to demonstrate it explicitly, but it is helpful 
to remember  that this must be the case. 
   
An extremely interesting issue is an existence and role of {\em non--classical}
saddle point solutions, which have $\Phi_q\neq 0$. Such solutions, if exist, have, 
in general, a  
non--zero action  and  may be loosely called {\em instantons}.   
Examples of  instantons include (thermal) escape from a quantum well 
\cite{Ioffe} and Wigner--Dyson statistics in metal grains \cite{Altland00}. 
Since the normalization $Z=1$ is entirely due to classical saddle points, 
instantons must have strictly zero contribution to the partition function, $Z$.
They may, however, contribute to correlation and response functions. 
The subject is  poorly developed (unlike instantons in the 
imaginary time) and requires much more attention.

\vskip 1cm 

8. {\em Semiclassical approximation.}
Let me concentrate now on a classical saddle point, Eqs.~(\ref{cl1}), (\ref{sp2}).
To include quantum effects in the semiclassical approximation one has to include 
fluctuations of the $\phi_q$ field near zero. To this end we keep terms up to the 
second order in $\phi_q$ in the action.  The resulting semiclassical action 
takes the following general form 
\begin{equation} 
S_{scl} = 2  \int\!\!\int\limits_{-\infty}^{\infty}\!\! dt dt'\, 
\left[ \bar\phi_q  [{\cal D}^{-1}]^K \phi_q
+ (\bar\phi_q {\cal O}^R[\phi_{cl}] \phi_{cl} + c.c.)\right] \, ;
                                                              \label{semi}
\end{equation}
$c.c.$ stays for the complex conjugation. From this point one may proceed in 
two directions:

(i) Since the action Eq.~(\ref{semi}) is Gaussian in $\phi_q$ one may integrate 
it out and end up with the theory of the single filed, $\phi_{cl}$. The corresponding 
action is   
\begin{equation} 
S_{scl}[\phi_{cl}] = 2  \int\!\!\int\limits_{-\infty}^{\infty}\!\! dt dt'\, 
 \bar\phi_{cl}  ({\cal O}^A[\bar\phi_{cl}] {\cal D}^A) [{\cal D}^{K}]^{-1} 
({\cal D}^R {\cal O}^R[\phi_{cl}]) \phi_{cl} \, .
                                                              \label{semi1}
\end{equation}
If the non--linearity in $\phi_{cl}$ is neglected, one ends up with the 
Gaussian theory governed by the inverse Keldysh Green function, 
$S\sim \bar\phi_{cl}   [{\cal D}^{K}]^{-1}  \phi_{cl}$ 
\footnote{If ${\cal D}^{R,A} = (0 \mp i\omega)^{-1}$, then  in the  
thermal equilibrium  (c.f. Eq.~(\ref{fdt})) , 
$[{\cal D}^{K}]^{-1} = {-i\over 2} \omega \mbox{tanh}{\beta\omega \over 2}$. 
In the $T=0$ limit one obtains the non--local quantum dissipative action,
$S\sim |\omega| |\phi_{cl}(\omega)|^2$, while in the high temperature regime 
the action is local $S\sim \beta |\dot \phi_{cl}(t)|^2$.}. 
In general, the action 
Eq.~(\ref{semi1}) is pretty complicated. If it contains only a finite number of the 
degrees of freedom and is short--ranged in time, 
one may use the transfer--matrix technique to write down 
the corresponding 
''Schr\"odinger'' equation. The result is the Focker--Planck equation.

(ii) One may perform the Hubbard--Stratonovich transformation with the auxiliary 
stochastic field $\xi(t)$ to decouple the quadratic term in Eq.~(\ref{semi}). The 
resulting action is  linear in $\phi_q$. Therefore the integration over
$\phi_q$ leads to the functional $\delta$--function. As a result one obtains 
stochastic Langevin equation
\footnote{One may worry about appearance of the Jacobean upon integration of 
the functional $\delta$--function over $\phi_{cl}$. It does not happen with our 
regularization of the functional integral (Jacobean is unity). The reason is 
exactly the same which guarantees $Z=1$ upon integration of the fluctuations 
around a classical saddle point.}   
\begin{equation}
{\cal O}^R[\phi_{cl}] \phi_{cl}(t) = \xi(t)
\, ,
                                             \label{lang}
\end{equation}
where $\xi(t)$ is a Gaussian noise with the correlator
\begin{equation}
\langle \xi(t) \bar\xi(t') \rangle = {i\over 2} [{\cal D}^{-1}]^K(t,t')
\, . 
                                             \label{noise}
\end{equation}
For the diffusive dynamics, $[{\cal D}^{R,A}]^{-1} = D\nabla^2 \mp i\omega$, 
in the thermal equilibrium 
$i [{\cal D}^{-1}]^K(\omega) = i \mbox{coth}{\beta\omega\over 2} 
([{\cal D}^{R}]^{-1}(\omega) - [{\cal D}^{A}]^{-1}(\omega)) \to 4T$ in the 
limit $\omega\ll T$. In this case one obtains the classical Langevin force with 
the correlator $\langle \xi(t) \bar\xi(t') \rangle = 2T\delta(t-t')$. The opposite 
limit is called sometimes the quantum Langevin equation. 

Notice that the above procedure may be reversed. Start from the classical Langevin
equation, then introduce functional $\delta$--function. Finally exponentiate it with 
the auxiliary field called $\phi_q$ and integrate out the Gaussian noise. Such procedure
is known in the literature as Martin--Siggia--Rose--DeDominics (MSRD)
\cite{MSR} technique. The result is the semi--classical Keldysh action,
Eq.~(\ref{semi}), in the high temperature limit. In the MSRD literature the field 
$\phi_{cl}$ is called $\phi$, while $\phi_q$ is usually called $\hat\phi$. 
One may wonder if one can form their symmetric and antisymmetric combinations 
$\phi_f = \phi+\hat\phi$ and  $\phi_b = \phi-\hat\phi$ to recover an 
''underlying quantum theory'' with the action $S[\phi_f]-S[\phi_b]$. In general 
it does not work since MSRD is only the semiclassical limit of Keldysh. Therefore 
terms   of the order $O(\hat\phi^3)$ may be ''missing''.

\vskip 1cm

9. {\em FDT and kinetic equation}.
Having understood the structure of the saddle points, we can construct a
perturbative expansion around a {\em classical} saddle point
\footnote{I am not aware about any attempt of doing the perturbation theory on top 
of a {\em non--classical} saddle point.}. 
Let say we are about to calculate the full Green function, defined as 
${\cal D}_{\alpha,\beta }(t,t')= 
-2i \langle \phi_\alpha(t)\bar\phi_\beta(t')\rangle$, where 
$\alpha,\beta = cl,q$.
This is done in a standard way by expanding the exponent of the non--Gaussian action 
and applying the Wick theorem according to Eq.~(\ref{Green}). 
One then rearranges the perturbation theory and defines the irreducible self--energy 
$\Sigma$ through the relation: 
\begin{equation}
{\cal D} = {\cal D}_0 + {\cal D}_0 \Sigma {\cal D}_0 + 
{\cal D}_0 \Sigma {\cal D}_0{\cal D}_0 \Sigma {\cal D}_0 +\ldots\, ,
                                             \label{self}
\end{equation}
where  multiplication is understood as the matrix one in the time and $cl-q$ spaces; 
${\cal D}_0$ is the bare propagator, Eq.~(\ref{Green}). 
The causality guarantees (this may be checked order by order)  that the 
self--energy matrix has the structure of the correlator 
(i.e. ${\cal D}_0^{-1}$; cf. Eq.~(\ref{action}))
\begin{equation} 
\Sigma_{\alpha,\beta}(t,t')= 
\left(\begin{array}{cc}
0   & \Sigma^{A}(t,t')  \\
\Sigma^{R}(t,t')  & \Sigma^K(t,t') 
\end{array}\right)
                                                             \label{sigma}
\end{equation}
Consider e.g. the second  term on the r.h.s. of Eq.~(\ref{self}), after a 
little algebra one finds
\begin{eqnarray}
&&\delta{\cal D}{cl,cl} =  {\cal D}_0^K \Sigma^A {\cal D}_0^A + 
{\cal D}_0^R \Sigma^R {\cal D}_0^K + {\cal D}_0^R \Sigma^K {\cal D}_0^A \, ;   
\nonumber \\
&&\delta{\cal D}{cl,q} =  {\cal D}_0^R \Sigma^R {\cal D}_0^R  \, ;   
\nonumber \\
&&\delta{\cal D}{q,cl} =  {\cal D}_0^A \Sigma^A {\cal D}_0^A  \, ;   
\nonumber \\
&&\delta{\cal D}{q,q}  = 0\, .
                                         \label{first}
\end{eqnarray}
As a result the $q-q$ component remains zero, while $cl-q$ and 
$q-cl$ remain retarded and advanced correspondingly. Indeed, 
product of triangular matrices is again a triangular matrix. 
Obviously, the same is true in all higher orders as well. 
Therefore 
the causality structure remains intact even in the presence of 
non--linearity (at least in perturbation theory).  
Eq.~(\ref{self}) may be rewritten as the matrix Dyson equation
\begin{equation}
\left({\cal D}_0^{-1} - \Sigma \right) {\cal D} = 1\, .
                                             \label{dyson}
\end{equation}
It is a simple matter to show that the $q-cl$ component of this equation
is nothing but the kinetic equation on the distribution function $F$.
Since this procedure is well documented in the literature 
\cite{Keldysh65,Landau,Rammer} we shall not describe it here
\footnote{The simplest exercise is the theory of the real bosonic field with 
the interaction $\lambda\phi^3 \to 3\lambda\phi_{cl}^2\phi_q + 
\lambda\phi_q^3$. In the second order the self--energy Fig.~\ref{phi32} 
is given by 
$\Sigma^{R,A} \sim \lambda^2{\cal D}_0^K*{\cal D}_0^{R,A}$ and 
$\Sigma^{K} \sim \lambda^2( {\cal D}_0^K*{\cal D}_0^{K} + 
{\cal D}_0^R*{\cal D}_0^{A} + {\cal D}_0^A*{\cal D}_0^{R})$, where 
the star denotes 
convolution in the energy space.  
\begin{figure}
\vglue 0cm
\hspace{0.01\hsize}
\epsfxsize=1.0\hsize
\epsffile{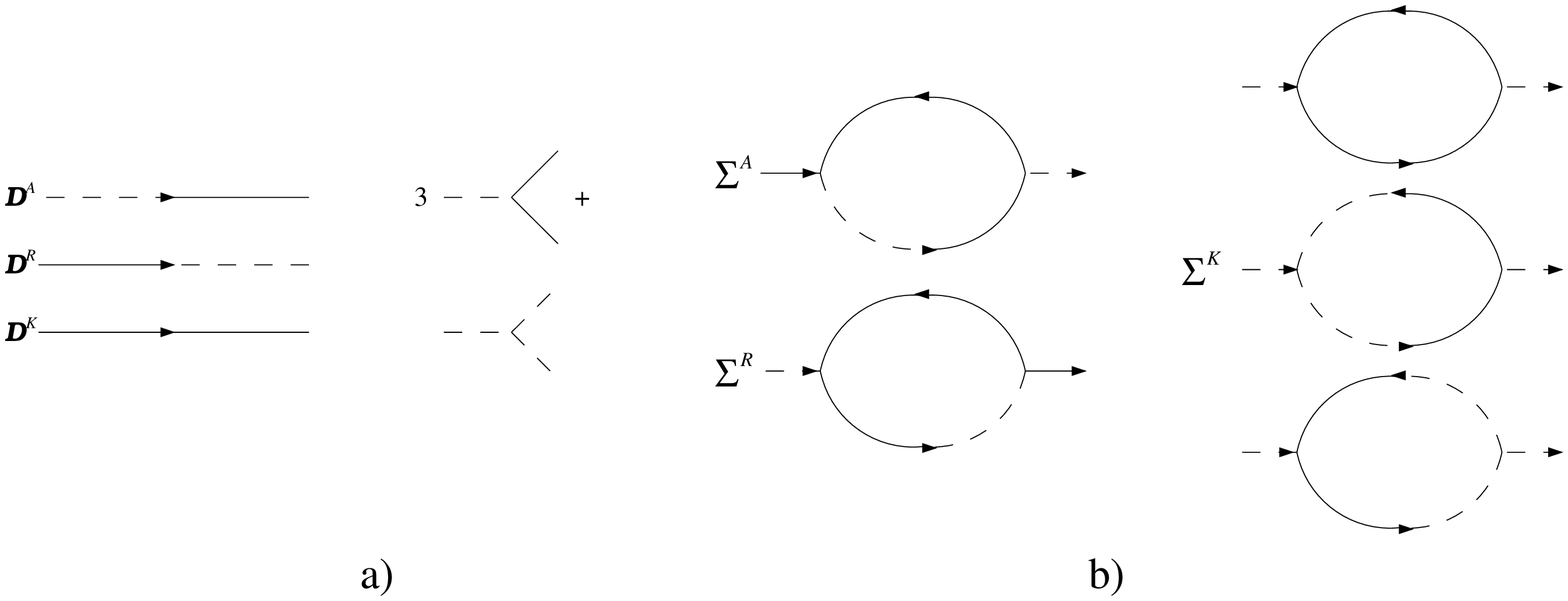}
\refstepcounter{figure} \label{phi32}
{\small FIG.\ \ref{phi32}
Example of the Keldysh diagrammatic for the $\phi^3$ 
theory. Fields $\phi_{cl}$ are denoted by full lines and 
$\phi_{q}$ by dashed lines. 
There are two vertices with relative coefficient $3$.
There is one retarded, one advanced and three Keldysh self--energy
diagrams 
  \par}
\end{figure}
Employing the ''magic'' identity
$\mbox{coth}\,\epsilon\, \mbox{coth}(\epsilon+\omega) - 1 = \mbox{coth}\, \omega
\left( \mbox{coth}\,\epsilon - \mbox{coth}(\epsilon+\omega) \right)$,
one may immediately see that if ${\cal D}_0$ satisfy FDT, Eq.~(\ref{fdt}),
then so does $\Sigma$, namely $\Sigma^K = \mbox{coth} {\beta\omega\over 2} 
(\Sigma^R-\Sigma^A)$. Then the Dyson equation~(\ref{dyson}) says that the 
exact propagator also satisfy FDT. The ''magic'' identity is directly 
responsible for the fact that the Planck distribution ($\mbox{coth} {\beta\omega\over 2}$) nullifies the collision integral. }.
Let me only mention that to obtain the textbook (semi--classical) 
kinetic  equation one has to assume that all external fields are slow 
on the scale of temperature. This assumption justifies the time--Wigner 
transform of the matrix $F$, which is a necessary step in the derivation of the 
kinetic equation. 
 
The other way to obtain the kinetic equation is to introduce time non--local  
field $Q(t,t')$, canonically conjugated to the composite operator 
$\phi(t)\bar\phi(t')$. A small amount of disorder makes $Q(t,t')$ a 
dynamic field which is governed by a certain effective action (this action may 
be obtained by integrating out the  fields $\phi$ and $\bar\phi$). 
The saddle point equation for this effective action $S_{eff}[Q]$ is 
the kinetic equation. As usually, the (semi) {\em classical} (kinetic) equation 
is obtained by variation over the {\em quantum} component of the $Q$--field.
For the details of this procedure see Ref.~\cite{Kamenev99}. 
One can possibly go beyond the saddle--point approximation by including the Gaussian 
fluctuations of the quantum component of $Q$ (pretty much in the way we did above).
The result of such procedure is probably  the Langevin--kinetic equation of 
Kogan and Shulman \cite{Kogan69}. To the best of my knowledge, it was not demonstrated explicitly yet.

\vskip 1cm

10. {\em Sources, external fields, response and correlation functions.} 
Up until now we have been busy with complicated representation of unity: $Z=1$. 
This is an interesting, but useless job. To make the technique work we need 
to calculate something more informative. This something is the generating 
function, which is obtained by adding the source fields to the action. It is 
convenient to discuss it in parallel with introduction of external classical 
fields (such as scalar or vector potentials). Let us add the source term
$S_{source} ={1\over 2}\int_{\cal C}\! dt\,V\bar\phi \phi$ to the action. 
The source field $V(t)$ is defined on the Keldysh contour, therefore 
it is rather a couple of fields $(V_f,V_b)$, residing on the forward and backward 
branches correspondingly. We decompose it now 
on the classical and quantum components exactly in the same way we did it with the 
$\phi$ field: $V_{cl}=(V_f+V_b)/2$ and $V_{q}=(V_f-V_b)/2$. The source 
action  takes the form 
\begin{equation}
S_{source} = \int\limits_{-\infty}^{\infty}\!\! dt\, 
\big[ V_{cl}(\bar\phi_{cl}\phi_q + \bar\phi_q\phi_{cl})  + 
V_q (\bar\phi_{cl}\phi_{cl} + \bar\phi_q\phi_{q})\big] \, . 
                                                       \label{source}
\end{equation}
Notice that the {\em quantum} source is coupled to the {\em classical} component of
the density (defined in a usual way as a half sum of the density 
on the forward and backward branches)  and vise--versa.
We define now the generating function as 
\begin{equation}
Z[V_{cl},V_q] = \left\langle e^{iS_{source}} \right\rangle \, . 
                                                       \label{generating}
\end{equation}
The classical component of the source field is nothing but classical external 
scalar potential, which is supposed to be the same on the two branches. The presence of 
such field  does not violate the fact that  the forward and the  
backward evolution brings the system back to the initial state. 
Therefore it does not change 
the normalization $Z=1$. We therefore arrive to the conclusion
\begin{equation}
Z[V_{cl},V_q=0] = 1\, . 
                                                       \label{knorm}
\end{equation}
One can also check this statement explicitly by e.g. expanding the action in 
powers of $V_{cl}$ and making contractions. All terms in this expansion 
vanish due to 
the fact that $\langle \phi_q \bar\phi_q \rangle = 0$. Eq.~(\ref{knorm}) is 
a close relative of Eq.~(\ref{causality}) and is just another manifestation of the 
causality. As a result, adding to the theory classical fields alone does not do any 
good -- we still work with unity. Therefore, one necessarily has  to introduce 
{\em quantum} sources (they change  sign by going from the forward to the 
backward branches of the contour). 
The presence of such source fields explicitly violates causality, and thus changes the 
partition (generating) function. On the other hand such fields usually do not  
have a physical meaning (see, however, the next section) and play only auxiliary role. 
In most cases one uses them only to generate observables (by appropriate differentiation) 
and then put them to  zero, restoring the causality structure of the action. 
Notice that the classical component, $V_{cl}$, does  {\em not} have to be 
taken to zero.   

Let us see how it works. Suppose we are interested in the average density of bosons 
at time $t$ in presence of a certain scalar potential $V_{cl}(t)$. Of course,
we are looking for the classical component of the density. According to 
Eqs.~(\ref{source}) and (\ref{generating}) it is given by 
\begin{equation}
n(t;V_{cl}) = -i\,{\delta \over \delta V_q(t)} Z[V_{cl},V_q] \Big|_{V_q=0}\, .
                                                       \label{density}
\end{equation}
To calculate this quantity one has to find first an appropriate (classical) saddle 
point of the action in presence of $V_{cl}$ (as was mentioned above this amounts  
to solving the kinetic equation) and then study quantum corrections. The problem
is simplified if the external field, $V_{cl}$, is weak in some sense. Then one may 
go away with the linear response
\begin{equation}
\chi^R(t,t') \equiv {\delta \over \delta V_{cl}(t')} n(t;V_{cl})\Big|_{V_{cl}=0} = 
-i\, {\delta^2 \over \delta V_{cl}(t') \delta V_q(t)} 
Z[V_{cl},V_q] \Big|_{V_q=V_{cl}=0}\, .
                                                       \label{linear}
\end{equation}
I put the subscript $R$, anticipating on the physical ground that the response 
function must be {\em retarded}. We shall demonstrate it momentarily. First, let us 
introduce for esthetical reasons the matrix
\begin{equation}
\chi_{\alpha,\beta}(t,t') \equiv 
-i\, {\delta^2 \over \delta V_{\beta}(t') \delta V_{\alpha}(t)} Z[V] \Big|_{V=0} = 
\left(\begin{array}{cc}
0   & \chi^A(t,t')  \\
\chi^R(t,t')  & \chi^K(t,t') 
\end{array}\right)\, .
                                                              \label{matrix}
\end{equation}
The fact that $\chi_{cl,cl} =0$ is obvious from Eq.~(\ref{knorm}). To evaluate 
the linear response matrix $\chi$ consider first the Gaussian action, 
Eq.~(\ref{action}). Adding the source term, Eq.~(\ref{source}), and integrating 
out the boson fields $\bar\phi$, $\phi$ one obtains
\begin{equation}
Z[V] = e^{-\Tr \ln (1 + {\cal D} \hat V)}\, ,
                                                              \label{trlog}
\end{equation}
where $\hat V = (V_{q}\sigma_0 + V_{cl}\sigma_1)/2$. Notice that the normalization 
is just right, since $Z[0] = 1$. One may now expand $\ln (1 + {\cal D} \hat V)$ to 
the second order in $\hat V$ and then differentiate twice. 
It is a straightforward matter 
to see that the causality structure of ${\cal D}$ imposes it also on $\chi$. 
E.g. $\chi^R(t,t') \sim \Tr \{{\cal D}\sigma_1{\cal D}\sigma_0\} = 
{\cal D}^R(t,t') {\cal D}^K(t',t) + {\cal D}^K(t,t') {\cal D}^A(t',t)$, which is 
obviously retarded. In presence of interactions, Eq.~(\ref{int}), one may 
perform Hubbard--Stratonovich transformation and then integrate out  
$\bar\phi$, $\phi$ fields. At this stage the causality structure of $\chi$ becomes 
apparent (it does not mean yet that one knows to evaluate $\chi$). 

The purpose of the above discussion was to demonstrate that even for the linear 
response problems (which may be always solved in Matsubara technique) the 
Keldysh formulation is natural and transparent. For one thing it allows one to 
overcome the analytical continuation procedure, which may happen to be a 
headache. The other advantage is that it gives both response and correlation 
functions in the single framework
\footnote{In the equilibrium response and correlation are connected by the 
FDT: $\chi^K(\omega) = \mbox{coth}{\beta\omega \over 2} 
(\chi^R(\omega)-\chi^A(\omega))$. However, out of the equilibrium they require 
separate calculations.}.

\vskip 1cm

11. {\em Counting statistics and counting fields.}
A very interesting application of the Keldysh formalism was suggested 
recently by L.~S.~Levitov and coworkers \cite{Levitov}. Suppose we are interested 
in a certain quantum mechanical observable $\hat O$ (for example, a total charge
which passed through a wire during a long time $t_f -t_i$). One may look for e.g. its 
(quantum) expectation value, $\langle \hat O \rangle$, or noise 
$\langle \hat O^2 \rangle - (\langle \hat O \rangle)^2$, or even higher order 
moments.  The most general quantity, however,
is the distribution function: $P(O)$ -- probability to observe the value $O$
in a single measurement of $\hat O$. If one knows $P(O)$, 
then $\langle \hat O^k \rangle = \int dO P(O) O^k$. 
Instead of discussing $P(O)$ it is convenient to consider its Fourier transform 
$F(\lambda) = \int dO P(0) e^{i\lambda O}$. The variable $\lambda$ - is called 
the counting field, and $P(O)$  (or $F(\lambda)$) is called the  counting statistics. 

From the above expressions one may deduce that 
\begin{equation}
F(\lambda) = \left\langle e^{i\lambda \hat O} \right\rangle\, .
                                                   \label{counting}
\end{equation}
The non--trivial fact is that the angular brackets in this expression must 
be understood as the averaging along the Keldysh contour. The prove of this 
statement requires a careful consideration of the time--ordering convention 
for the higher moments of $\hat O$. We refer a reader to the original works 
\cite{Levitov,Nazarov} for more details. If the Keldysh nature of the averaging in
Eq.~(\ref{counting}) is accepted, it should be obvious by now that the 
counting field $\lambda$ is a purely  {\em quantum} field. Indeed, $\lambda$ was 
introduced to generate moments of $\hat O$ upon differentiation. As we have seen
in the previous section -- it is the quantum component of the source field 
(having different sign on forward and backward branches (!)), 
which serves for this purpose. As a result, Eq.~(\ref{counting}) suggests that the 
Keldysh action of the system at hand must be modified as
\begin{equation}
S^{\lambda}[\phi] = S[\phi] + \int\limits_{-\infty}^{\infty} \! dt\, 
\lambda_q(t)  \hat o_{cl}(t)  \, .
                                                   \label{modif}
\end{equation}
We put $\hat O =\int_{t_i}^{t_f} \! dt\, \hat o_{cl}(t)$, where the 
classical observable $\hat o_{cl}(t)$ is expressed through the  fields $\phi(t)$.
We have also put $\lambda_q(t) \equiv \lambda \theta(t-t_i)\theta(t_f-t)$ -- thus the name --  {\em counting} field.

The novel feature is that the quantum source, $\lambda_q$, should {\em not} be taken 
to zero in the end of calculations. It rather remains finite and possibly even 
large to allow the inverse Fourier transform of $F(\lambda)$. The presence of 
the finite quantum source in the action spoils immediately the causality structure 
of the theory. For example, $S^{\lambda}[\phi_q = 0]\sim \lambda \neq 0$. 
As a result,  
$\Phi_q = 0$ is {\em not}  a  saddle point solution any more. There are 
no classical saddle points in the theory ! A new non--classical saddle point 
has in general a non--zero  action $S_{sp} = S_{sp}(\lambda) \neq 0$. 
Neglecting fluctuation effects, one finds for the generating function of 
the irreducible moments: $-i\ln F(\lambda) = S_{sp}(\lambda)$.
Such program (including fluctuations) was realized by Yu.~V.~Nazarov 
\cite{Nazarov} for the counting statistics of charge passing through a diffusive 
wire. See also related works on counting statistics  of adiabatic pumps 
\cite{Andreev00}. 

To conclude this section I want to mention a striking similarity between counting 
statistics in Keldysh theory and statistics of rare events in supersymmetric 
technique \cite{Khmelnitskii}. To my opinion this subject is far from being closed.

\vskip 1cm

12. {\em Keldysh technique in disordered systems.} 
Due to the presence  of the intrinsic normalization condition, $Z=1$, the 
Keldysh technique is ideally suited for treatment of disordered systems,
such as spin glasses or dirty metals. Notice that the response and correlation 
functions were defined by differentiation of $Z[V]$, Eq.~(\ref{matrix}), rather 
than more familiar $\ln Z[V]$ (since $Z[V_q=0] = 1$ this is completely equivalent). 
Let me also mention again that the normalization constant, 
$\Tr\{\rho_0 \}$, introduced in Eq.~(\ref{e1}), is disorder--{\em independent}. 
The point is that disorder is  switched on 
(and off) on forward (backward) parts of the contour  after (before)
$t=-\infty$. As a result, the ''partition'' or generating function $Z$ (rather 
than its logarithm) is the proper object to be averaged over the  quenched disorder. 
This circumvent the need to introduce replica \cite{Edwards}
(or supersymmetry \cite{Efetov}) to take care of the logarithm. 
This was first realized in the context of classical spin glasses (MSRD version of 
Keldysh) by H.~Sompolinsky \cite{Sompolinsky} and substantially 
advanced later \cite{Feigelman}.   
In the context of  electrons in the random potential the corresponding theory 
(Keldysh non--linear $\sigma$--model) was developed in 
Refs.~\cite{Kamenev99,Shamon99,Larkin00}. 

Here I restrict myself to a few remarks only. On the perturbative level the 
replica and supersymmetry methods are supposed to eliminate  ''parasitic''
diagrams.  Such decoupled closed loop diagrams are generated if one erroneously 
averages $Z$ instead of $\ln Z$. The replica trick first prescribes  factor $n$ 
(number of replica) to such loops and then eliminates them by taking the limit 
$n\to 0$. In the supersymmetry bosonic loops exactly cancel fermionic ones.
The Keldysh analog of such cancellation mechanism is given by 
Eqs.~(\ref{traces}) and (\ref{traces1}). Indeed, the closed loops coupled to the 
rest of a diagram by the static disorder  carry either only retarded or only 
advanced Green functions
\footnote{Indeed, the static disorder $U_{dis}$ enters the Keldysh matrix action
as $U_{dis}\sigma_1$. Integrating out of the $\phi$ fields
one ends up with $\Tr\ln(1+{\cal D}\sigma_1U_{dis})$, cf. Eq.~(\ref{trlog}).
Expanding the logarithm one finds terms of the form 
$\Tr\{{\cal D}\sigma_1\ldots {\cal D}\sigma_1\}$, the later are obviously 
$\Tr\{{\cal D}^R\ldots {\cal D}^R\} + \Tr\{{\cal D}^A\ldots {\cal D}^A\} = 0$.}. 
The later vanish according to Eqs.~(\ref{traces})  (\ref{traces1}).
Therefore it is not surprising that all {\em perturbative} results of the 
non--linear $\sigma$--model (including electron--electron interactions)
are correctly and elegantly reproduced in the Keldysh version 
\cite{Kamenev99,Shamon99}. 

Whether or not the Keldysh  $\sigma$--model contains  the 
{\em non--perturbative} effects (such as instantons)  -- is a hundred dollar 
question. I believe that the answer is positive. At least, it is possible to
obtain the non--perturbative (though only asymptotic) part of the level statistics 
via the Keldysh $\sigma$--model \cite{Altland00}. 
To this end, one has to analyze carefully non--classical (having a 
non--zero action) saddle points on the massless manifold of the $\sigma$--model.
In any case, a lot to be done before the Keldysh technique will become 
a reliable tool for the non--perturbative calculations.


\vskip 1cm

\section{Doi--Peliti technique for classical systems}

1. {\em A bit of history.} The technique was invented by Doi in 1976 
\cite{Doi} and later rediscovered by Peliti \cite{Peliti}, 
and others \cite{Matis,GS}. J.~Cardy and B.~P.~Lee were probably most persistent in applying 
the technique to various non--equilibrium statistical problems 
\cite{Cardy}. These notes are partially based on lectures of J.~Cardy 
\cite{Cardy1}.

\vskip 1cm

2. {\em Motivation.} We are talking about extremely broad and diverse 
types of problems
whose applicability ranges from  
epidemiology and dynamics of biological populations 
to models of chemical reactions and glasses. The problems at hand are usually 
determined on some $d$--dimensional lattice. They are defined by specifying the 
following entities:
\begin{itemize}
\item The agents, usually named as $A, B,\ldots$ These may be molecules, 
animal species, stock options, vehicles, {\em etc}.
\item Rules of agent's motion on the lattice. It may be e.g. random walk, or 
directed ballistic motion. 
\item Rules of agent's interactions. E.g. the simplest one is the binary 
annihilation: if two agents meet at the same lattice site they annihilate with 
probability $\mu$. The short way to write is 
\begin{equation}
A + A \stackrel{\mu}{\rightarrow}0\, .
                                              \label{ba}
\end{equation}
This model may, for example, describe a simple chemical reaction, where 
two identical molecules form a stable aggregate which segregates from the solution. 
An other famous example is the Lotka--Volterra system \cite{lotka,oded} 
\begin{eqnarray}
&& A \stackrel{\mu} {\rightarrow} 0\, ; \nonumber  \\
&& B \stackrel{\sigma}{\rightarrow} 2B\, ; \label{lv} \\
&& A + B \stackrel{\lambda}{\rightarrow} 2A\, , \nonumber  
\end{eqnarray}
where $A$ designate ``foxes'' and $B$ --- ``rabbits''. Foxes die with probability 
$\mu$; rabbits proliferate with probability $\sigma$, if a fox and a rabbit meet --- the rabbit is eaten and another fox is produced  with probability $\lambda$.

\end{itemize}

\vskip 1cm

3. {\em Mean--field.} The simplest way to treat systems described by 
Eqs.~(\ref{ba}) or (\ref{lv}) is to employ the {\em mean field} equations.
To this end one introduces the mean density of agents, say $A$, at a 
lattice site $r$ at time $t$ -- $n_A(r,t)$. 
For the system Eq.~(\ref{ba}) with random 
walk motion (ARW -- annihilating random walk) 
the mean field equation for the density is 
\begin{equation}
\frac{\partial n_A}{\partial t} =
D \nabla^2_r n_A  -\mu n_A^2\, ,
                                              \label{bmf}
\end{equation}
where $D$ is the diffusion rate. Derivation of the mean field equation is 
self--evident from the definition of the model, Eq.~(\ref{ba}). We shall see 
later, that from the perspective of  Doi--Peliti technique the mean--field 
is nothing but equation for a {\em classical} saddle point. A reader is advised 
to write down and investigate the mean--field equations for the Lotka--Volterra
system, Eq.~(\ref{lv}). 

The mean--field ignores fluctuations. It works reasonably well, when the number 
of agents at each site is large. In the opposite limit where the  
average number of agents on a site is less then unity, 
the system's behavior may be solely determined by 
fluctuation effects (especially in low spatial dimensionality). For example, 
Eq.~(\ref{bmf}) predicts $n_A \sim t^{-1}$ at $t\to \infty$, while the correct 
result \cite{Cardy,Cardy1} is $n_A \sim t^{-d/2}$ for $d<2$ and 
$n_A \sim t^{-1}$ for $d >2$. The behavior is clearly very different from the 
mean--field prediction for dimensionality below the critical one, $d_c =2$. 
A traditional approximate way to handle fluctuations (which also stops short 
to obtain the above mentioned results) is by means of the Focker--Planck 
or Langevin equations. I shall comment below on the approximations 
involved in their derivation.

\vskip 1cm

4. {\em Master equation.} The formally exact way to handle the problem is the 
master equation. One first defines a many--body microscopic configuration $\alpha$,
determined by the (integer) number of agents at each lattice cite
\begin{equation}
\alpha = \{n_1, n_2,\ldots n_M \} \, ,
                                              \label{alpha}
\end{equation}
where $M=L^d$ is the total number of cites in the $d$--dimensional lattice.
Next one defines $P(\alpha, t)$ -- probability to find the system in the 
many--body configuration $\alpha$ at time $t$. The master equation has the form
\begin{equation}
\frac{d}{d t} P(\alpha, t) = 
\sum\limits_{\beta} {\cal R}_{\beta\to\alpha} P(\beta,t) -
\sum\limits_{\beta} {\cal R}_{\alpha\to\beta} P(\alpha,t)\, ,
                                              \label{me}
\end{equation}
where the sums run over all many--body configurations $\beta$. The first term 
on the r.h.s. describes ``in'', while the second one ``out'' processes. The 
coefficients ${\cal R}_{\beta\to\alpha}$ and ${\cal R}_{\alpha\to\beta}$ are the 
transition rates. A particular case when,
\begin{equation} 
\frac{ {\cal R}_{\beta\to\alpha}  }
{ {\cal R}_{\alpha\to\beta} } =
\frac{ W(\beta)}{W(\alpha)}\, , 
                                              \label{db}
\end{equation}
where $W$ is a certain positive function uniquely defined for each state, is 
said to satisfy the {\em detailed balance} condition.  

As an example let us consider the binary annihilation on a single site. The 
microscopic state is specified by a positive integer number of agents,  
$\alpha = \{n\}$. According to Eq.~(\ref{ba}) the master equation takes the form
\begin{equation}
\frac{d}{d t} P(n, t) = 
\mu\,  \frac{(n+2)(n+1)}{2}\, P(n+2,t) -
\mu\, \frac{n(n-1)}{2}\,  P(n,t)\, .
                                              \label{bame}
\end{equation}
The only non--zero rates are  ${\cal R}_{n+2\to n}$ and 
${\cal R}_{n\to n-2}$, which are proportional to the number of possible 
{\em pairs} in the corresponding state. These rates do not satisfy the detailed balance. 

\vskip 1cm

5. {\em Quantization.} The basic idea \cite{Doi} is to draw the analogy between 
the master equation and the many--body Schr\"odinger equation. Indeed, 
both equations are linear in $P$ (or $\Psi$) and first order in $\partial_t$. 
To make the analogy explicit one introduces the ``second quantization''.  
Define the ket--vector
$|n\rangle$ as the microscopic state with $n$--agents. Let us also define vector
\begin{equation}
|\Psi(t)\rangle \equiv 
\sum\limits_{n=0}^{\infty} P(n,t)\, |n\rangle \, .
                                              \label{psi}
\end{equation}
Note that the weight, $P$, is probability rather  than  the amplitude. 
In these notations the master equation takes the form
\begin{equation}
\frac{d}{dt}\, |\Psi(t)\rangle = 
\sum\limits_{n=0}^{\infty} \frac{d P(n,t)}{dt}\, |n\rangle = 
                                           \label{sch} 
\frac{\mu}{2}\, \sum\limits_{n=0}^{\infty} \Big[ 
(n+2)(n+1) P(n+2,t) - n(n-1) P(n,t) \Big] |n\rangle \, .  
\end{equation}
The aim now is to write the r.h.s. of this expression as 
$-\hat H |\Psi(t)\rangle$, where $\hat H$ is the ``Hamiltonian'' operator. 
To this end we introduce the creation and annihilation operators:
\begin{eqnarray}
a^{\dagger}|n\rangle  &=&  |n+1 \rangle \, ;    \label{a} \\
a|n\rangle  &=& n |n-1 \rangle   \, .
\end{eqnarray}
As a byproduct, one has $a|0\rangle = 0$. One may immediately check that 
such operators are ``bosonic''
\footnote{ In problems where not more than one agent is allowed on a given site,
(e.g. traffic flow models) one may probably employ ``fermions''. I am not aware, 
however, of  such attempts.}  
\begin{equation}
[a,a^{\dagger}] = 1 \, .
                                            \label{commutation}
\end{equation}
As for any pair of operators satisfying Eq.~(\ref{commutation}) one may prove 
the identity
\begin{equation}
e^a f(a,a^{\dagger}) =  f(a,a^{\dagger} + 1) e^a \, ,
                                            \label{identity}
\end{equation}
where $f$ is an arbitrary operator--value function. 
With these definitions one may  check that the Hamiltonian, we are 
looking for, has the form 
\begin{equation}
\hat H = {\mu \over 2}  \big(
(a^{\dagger})^2 a^2 - a^2 \big) \, ,
                                            \label{H}
\end{equation}
where the first term on the r.h.s. is the ``out''  and the second one 
is ``in'' term. Generalization for the many site problem with the random walk
(ARW) is straightforward, the resulting Hamiltonian is
\begin{equation}
\hat H = D \sum\limits_{<ij>} 
(a^{\dagger}_i - a^{\dagger}_j)(a_i - a_j) + 
{\mu \over 2}\, \sum\limits_j \big(
(a^{\dagger}_j)^2 - 1 \big) a_j^2    \, ,
                                            \label{H1}
\end{equation}
where $<ij>$ are nearest neighbors. 

\vskip 1cm

6. {\em Observables.} So far we succeed to put the master equation into 
the Schr\"odinger form
\begin{equation}
\frac{d}{dt}\, |\Psi(t)\rangle = -\hat H \, |\Psi(t)\rangle \, .
                                                  \label{schrodinger}
\end{equation}
The differences with the true Schr\"odinger equation are: 
\begin{itemize}
\item There is no imaginary unity, meaning that we are dealing with the 
imaginary time Schr\"odinger equation.
\item $\hat H$ is, in general, non--Hermitian (unless the detailed balance 
condition is satisfied).  
\item The most important: the mean value of some observable $\hat O$, is 
{\em not} $\langle \Psi(t)| \hat O |\Psi(t) \rangle$. Indeed, such combination
is quadratic in $P$, while the mean value should be linear in probability, $P$.
\end{itemize} 
To obtain the proper mean value one defines the bra--coherent state
\begin{equation}
\langle \Psi_0 | \equiv \langle 0|\,  e^a \, .
                                            \label{coherent}
\end{equation}
This state has the special property: $\langle \Psi_0 | n\rangle = 1$ for 
any $n$. This fact may be checked using Eq.~(\ref{identity}). Employing this 
coherent state, one may show that the mean value of an observable may be written
as 
\begin{equation}
\langle \hat O \rangle = \langle \Psi_0| \hat O |\Psi(t) \rangle  \, .
                                            \label{mean}
\end{equation}
Indeed, let us consider density as an example,  $\hat O = a^{\dagger} a$
\begin{equation}
\langle n(t) \rangle = \langle \Psi_0| a^{\dagger} a |\Psi(t) \rangle  =
\langle \Psi_0| a |\Psi(t) \rangle    
                                                         \label{meand}
=\langle 0|e^a  a^{\dagger} a\sum\limits_n P(n,t) |n \rangle=
\sum\limits_n n P(n,t) \, ,  
\end{equation} 
in agreement with the expectations based on the meaning of  $P(n,t)$. 
The second equality on the r.h.s. is very important; it is an immediate consequence of the coherent state being the left eigenstate of the creation operator, $\langle \Psi_0 | a^{\dagger} = \langle \Psi_0 | $.

\vskip 1cm

7. {\em Normalization.} To have a consistent theory, one has to worry that the 
total probability to find the system in some state is unity
\begin{equation}
1 = \sum\limits_n P(n,t) = \langle \Psi_0|\Psi(t) \rangle  \, .
                                            \label{norm}
\end{equation}
The last equality is just the mean value of the unit operator. One may solve 
formally the Schr\"odinger equation and write 
$|\Psi(t) \rangle = \exp\{-\hat H(a^{\dagger}, a) t\} |\Psi(0) \rangle$. 
Eq.~(\ref{norm}) is satisfied iff $\langle \Psi_0| \hat H(a^{\dagger}, a) =0 $.
Since the coherent state is an eigenstate of the creation operator,
$\langle \Psi_0| a^{\dagger} = \langle \Psi_0|$, one arrives at the conclusion 
that any legitimate Hamiltonian must obey
\begin{equation}
\hat H(a^{\dagger} = 1, a) = 0 \, .
                                            \label{norm1}
\end{equation}
E.g. the Hamiltonian of ARW, Eq.~(\ref{H1}), indeed satisfy this condition. 
To introduce the functional integral representation we'll need to have the normal ordering of creation and annihilation operators.
Therefore it is a good idea to 
commute $e^a$ through the Hamiltonian employing Eq.~(\ref{identity}).  
\begin{equation}
\langle 0|e^a  e^{-\hat H(a^{\dagger}, a) t} |\Psi(0) \rangle=
\langle 0|e^{-\hat H(a^{\dagger}+1, a) t} e^a |\Psi(0) \rangle\, .
                                                 \label{comutation}
\end{equation} 
If an observable is considered, the same commutation may be performed 
provided that the observable is written in terms of annihilation operators 
only. An example, of this procedure is provided by the second equality 
in Eq.~(\ref{meand}). We shall redefine from now on the Hamiltonian 
as $H(a^{\dagger}, a) \equiv \hat H(a^{\dagger}+1, a)$. In terms of the new 
Hamiltonian the fundamental normalization condition, Eq.~(\ref{norm}), is 
satisfied iff
\begin{equation}
H(a^{\dagger} = 0, a) = 0 \, .
                                            \label{norm2}
\end{equation}
We shall see momentarily that this condition is the direct analog of the Keldysh
normalization condition, Eq.~(\ref{causality}).

\vskip 1cm

8. {\em Field theory.}  Consider the ``partition function'' of the theory
\begin{equation}
Z\equiv \langle 0|  e^{- H(a^{\dagger}, a) t} |\tilde\Psi(0) \rangle \, ,
                                                 \label{partition}
\end{equation} 
where $|\tilde\Psi(0) \rangle = e^a\,|\Psi(0) \rangle$ is the renormalized 
initial microscopic state, its precise form is not very important for 
our further discussion. Due to normalization condition and according to 
Eqs.~(\ref{norm}), (\ref{comutation}), $Z=1$. We divide now the time interval  
$[0,t]$ onto $N\to \infty$ slices and introduce the coherent state resolution 
of unity at each time slice. As a result, one obtains in the standard way 
\cite{Negele} 
\begin{equation}
Z  =  
\int\! {\cal D}\phi^* \phi\,  \exp\{-S[\phi^*(t),\phi(t)] \} \, ,
                                                 \label{functional}
\end{equation} 
where 
\begin{equation}  
S[\phi^*(t),\phi(t)] = \int\limits_0^t \!\! dt\, 
\Big[
\phi^* \partial_t\phi + H(\phi^*,\phi) 
\Big] \, . 
                                                 \label{dpaction}
\end{equation} 
For example, in the case of ARW one has 
\begin{equation}  
S[\phi^*(t,r),\phi(t,r)] = \int\limits_0^t \!\! dt\, \int\!\!dr\, 
\Big[
\phi^* (\partial_t - D \nabla^2_r) \phi + 
{\mu\over 2} \left( 
(\phi^*)^2 \phi^2 +2 \phi^* \phi^2
\right)   
\Big] \, . 
                                                 \label{ARWaction}
\end{equation} 

\vskip 1cm

9. {\em Analogy with the Keldysh technique.} Notice that both in Keldysh and 
Doi--Peliti techniques we deal with the ``partition function'' $Z=1$. 
(No matter that the reasons  for this normalization are very different:
closed time contour in Keldysh versus conservation of probability in 
Doi--Peliti.) In both cases we succeed to write the partition function 
as the coherent state functional integral, cf. Eq.~(\ref{e2}) and 
Eq.~(\ref{functional}). To this end in both cases we had to {\em double} 
number of degrees of freedom: classical and quantum in Keldysh versus  
$\phi$ and $\phi^*$ in Doi--Peliti
\footnote{Notice that since we are interested in {\em real} density of 
agents, one may naively expect to obtain a field theory of the single 
real bosonic field. We have to deal with both  $\phi$ and $\phi^*$, however,
despite of the fact that the object is purely classic and the phase has no 
evident meaning.}. 
The normalization requires the strong condition on an acceptable action:
Eq.~(\ref{causality}) in Keldysh versus 
\begin{equation}  
S[\phi^* = 0, \phi ] = 0\,  
                                                 \label{dpnorm}
\end{equation} 
in Doi--Peliti. This equation is a direct consequence of Eqs.~(\ref{norm2}) and 
(\ref{dpaction}). Comparing Eqs.~(\ref{causality}) and (\ref{dpnorm}) one may 
suspect the analogy 
\begin{equation}  
\phi^* \sim \phi_q \, ; \hskip 2 cm \phi \sim \phi_{cl} \, .
                                                 \label{analog}
\end{equation} 
We shall see below that this analogy indeed works for all intends and purposes.
Let us rewrite the quadratic part of the action of the ARW model in 
the following form 
\begin{equation} 
S[\phi^*,\phi] = {1\over 2} \int\!\! dt \, 
(\phi,\phi^*) 
\left(\begin{array}{cc}
0   & \overbrace{-\partial_t - D\nabla^2_r}^A \\
\underbrace{\partial_t - D\nabla^2_r}_R & \underbrace{ X }_K 
\end{array}\right)
\left(\begin{array}{c}
\phi \\ \phi^*
\end{array}\right) \, .
                                                              \label{quadratic}
\end{equation}
Here $R$ and $A$ designate retarded and advanced linear classical operators.
Symbol $K$ stays for the Keldysh component of the correlator. In our simplistic 
example it is absent, $X=0$. This fact is not generic by any means:
in more complicated models e.g. Lotka--Volterra there is non--zero Keldysh 
component. Moreover, even if it is zero in the bare action -- it is usually generated 
in the process of renormalization. On the other hand, the fact that the 
$\phi-\phi$ element is zero is not accidental. This zero is direct 
a consequence of Eq.~(\ref{dpnorm}) and it is protected against any renormalization.

Once the structure of the quadratic action is understood, one may 
write down propagators to build the perturbation theory (cf. Eq.~(\ref{Green}) )
\begin{eqnarray}
\langle \phi(t) \phi^*(t')\rangle &=& 
{\cal D}^R (t,t') \sim (D q^2 - i\omega)^{-1} \, ;
\nonumber \\
\langle \phi^*(t) \phi(t')\rangle &=& 
{\cal D}^A(t,t')  \sim (D q^2 + i\omega)^{-1}\, ;
\nonumber \\
\langle \phi^*(t) \phi^*(t')\rangle &=& 0  \, ;
\nonumber \\ 
\langle \phi(t) \phi(t')\rangle &=& 
{\cal D}^K (t,t') \, . 
                                         \label{dpcorr}
\end{eqnarray}
As in the Keldysh case, ${\cal D}^R (t,t) = {\cal D}^A (t,t) = 0$. 
It is exactly the same causality structure, which is responsible for the conservation of probability (normalization) and the entire internal consistency of the theory.   
The analogy between the two techniques goes actually much further 
than the perturbative expansion. 

\vskip 1cm 

11. {\em Saddle point equations.}
Exactly as in the case of Keldysh, the saddle point equation
\begin{equation}
{\delta S\over \delta \phi } = 0\, 
                                             \label{dpsp1}
\end{equation}
may be always solved by 
\begin{equation}
\Phi^* = 0 \, .
                                             \label{dpcl1}
\end{equation}
The saddle point solutions are denoted by capital $\Phi$. 
Under the condition Eq.~(\ref{dpcl1}) the 
second saddle point equation takes the form
\begin{equation}
{\delta S\over \delta \phi^* } = 
{\cal O}^R[\Phi] \Phi =0
\, ,
                                             \label{dpsp2}
\end{equation}
where ${\cal O}^R[\Phi] $ is  the retarded operator. 
In the example of ARW, Eq.~(\ref{ARWaction}) 
\begin{equation}
{\cal O}^R[\Phi]  = \partial_t - D \nabla^2_r +\mu \Phi  \, 
                                             \label{mf1}
\end{equation}
and the saddle point equation~(\ref{dpsp2}) is nothing but the mean--field 
equation, Eq.~(\ref{bmf}). 
Notice, that the classical (mean--field) equation of motion is  a 
result of   variation
over the ``quantum'' ($\phi^*$)   component 
(in the limit where the later is zero).
In view of Eqs.~(\ref{dpnorm}) and (\ref{dpcl1}), the action 
on any classical solution is zero. 

The existence and role of ``non--classical'' saddle points 
(with $\Phi^* \neq 0$ and $S_{sp}\neq 0$) in the context of 
Doi--Peliti technique, to the best 
of my knowledge, was not yet investigated.

\vskip 1cm 

12. {\em Semiclassical approximation.}
To include fluctuation effects in the semiclassical approximation one 
keeps terms up to the second order in $\phi^*$ in the action. 
For the ARW problem the entire action is semiclassical (this is  not  generic  by any means). Let us rewrite it in the following
generic semiclassical form 
\begin{equation} 
S_{scl} =  \int dt \, 
\left[{1\over 2} \phi^*  [{\cal D}^{-1}]^K \phi^*
+ \phi^* {\cal O}^R[\phi] \phi \right] \, .
                                                              \label{dpsemi}
\end{equation}
For the ARW problem $ [{\cal D}^{-1}]^K = \mu \phi^2 >0 $. 
Positive definiteness of this quantity is necessary for the convergence of the $\phi^*$ integral. 
From this point one may again proceed in two directions:

(i) Since the action Eq.~(\ref{dpsemi}) is Gaussian in $\phi^*$ one may integrate 
it out and end up with the theory of the single filed, $\phi$. The corresponding 
action is   practically identical to Eq.~(\ref{semi1}). 
Employing the transfer--matrix method, one arrives at  
the Focker--Planck equation.

(ii) One may perform the Hubbard--Stratonovich transformation with the auxiliary 
stochastic field $\xi(t)$ to decouple the term quadratic in $\phi^*$. 
Subsequent integration over
$\phi^*$ results in  the functional $\delta$--function. This way  one obtains 
stochastic Langevin equation
\begin{equation}
{\cal O}^R[\phi] \phi(t) = i \xi(t)
\, ,
                                             \label{dplang}
\end{equation}
where $\xi(t)$ is  Gaussian noise with the correlator
\begin{equation}
\langle \xi^2(t)   \rangle =  [{\cal D}^{-1}]^K   \, . 
                                             \label{dpnoise}
\end{equation}
Some studies start from postulating Langevin (or Focker--Planck) 
dynamics for problems at hand. We note that, this approach is not exact, 
but rather a semiclassical approximation. In some 
cases, like the ARW, the semiclassical approximation and hence the 
Langevin dynamics is indeed exact. However, even in such cases the 
correlator of noise is far from being trivial. Moreover, the noise
term in Langevin equation, Eq.~(\ref{dplang}), contrary to any intuition,  
happens to be pure imaginary. It is hard to imagine how such dynamics could 
be postulated from the outset. 

\vskip 1cm 

13. {\em Sources.}
To calculate an observable, e.g. mean density, one need to introduce sources.
As was mentioned above, mean value of any observable may be expressed through 
annihilation operators only, c.f. the second equality in Eq.~(\ref{meand}). 
In the functional integral language it means that pre--exponential factors 
at hand are given in terms of the ``classical'' field, $\phi$, only. Therefore 
the source term, one needs, has the form $\int dt V^*(t)\phi(t)$ and the 
generating functional is given by 
\begin{equation}
Z[V^*] = \Big\langle e^{\int dt V^*(t)\phi(t) } \Big\rangle \, ,
                                         \label{dpgenerating}
\end{equation}
where the angular brackets stay for averaging with the action, 
Eq.~(\ref{dpaction}). With this generating functional one may obtain, e.g., 
the mean density of agents
\begin{equation}
\langle n(t) \rangle = 
\langle \Psi_0| a |\Psi(t) \rangle  = 
\int\! {\cal D}\phi^* \phi\,  e^{-S[\phi^*(t),\phi(t)]} \phi(t) = 
\frac{\delta}{\delta V^*(t)} Z[V^*] \Big|_{V^* = 0}  \, .
                                                         \label{fmeand}
\end{equation} 
Other observables are evaluated in the similar fashion. Note that as in 
the Keldysh technique, observables are given in terms of the ``classical'' 
field, $\phi$, 
while the generating function is functional of the fictions ``quantum'' source,
$V^*(t)$, which is to be taken to zero after differentiation.

\vskip 1cm

14. {\em Short summary.} The aim of this chapter is not to give a 
detailed account of a specific calculation employing the Doi--Peliti technique.
Lectures and papers of J.~Cardy {\em et al.} \cite{Cardy1,Cardy} are highly
recommended  for this purpose. My intention was to highlight the similarities 
and common concepts with the Keldysh techniques. The hope is to facilitate 
intermixing of the two largely non--intercepting communities. Let me stress 
once again common points of the two techniques:
\begin{itemize}
\item Internal fundamental {\em normalization}, $Z=1$.
\item Doubling of degrees of freedom: ``{\em classical--quantum}''.
\item Causality: {\em retarded--advanced} structure of the action and the propagators.
\item Existence of the {\em classical} extrema: $\Phi_q=0\,;\,\, 
{\cal O}^R \Phi_{cl} = 0$.
\item Langevin and Focker--Planck approaches as the  {\em semiclassical} 
approximation --- second order in $\phi_q$.
\item Generating function of {\em classical} observables is  
functional of the {\em quantum} source. 
\end{itemize} 

I am indebted to V.~Lebedev, A.~Andreev, A.~Altland and V.~Elgart for teaching me various 
aspects of the techniques and to L.~V.~Keldysh, M.~Doi and L.~Peliti for inventing them. 
This research was supported in part by the BSF grant N  9800338.

\end{document}